\documentclass[11pt]{article}
\usepackage{hyperref}
\usepackage{fullpage}

\pdfoutput=1

\begin{document}

\title{Non-coalescence of oppositely charged drops}

\author{W. D. Ristenpart$^{1}$, J. C. Bird$^{2}$, A. Belmonte$^{3}$,\\ F. Dollar$^{2}$ \& H. A. Stone$^{2}$\\
\vspace{6pt}\\
$^1$ Dept. Chem. Engr. \& Mater. Sci., Univ. California Davis\\
$^2$ School Engr. App. Sci., Harvard Univ.\\
$^3$Pritchard Lab, Dept. Math., Penn. State Univ.}

\maketitle

\begin{abstract}
Oppositely charged drops have long been assumed to experience an attractive force that favors their coalescence. In this fluid dynamics video we
demonstrate the existence of a critical field strength above which oppositely charged drops do not coalesce. We observe that appropriately
positioned and oppositely charged drops migrate towards one another in an applied electric field; but whereas the drops coalesce as expected at
low field strengths, they are repelled from one another after contact at higher field strengths. Qualitatively, the drops appear to `bounce' off
one another. We directly image the transient formation of a meniscus bridge between the bouncing drops.\\
\vspace{6pt}\\For details please see Ristenpart et al., \emph{Nature} \textbf{461}, 377-380 (2009).
\end{abstract}

\section*{Movie description}
The video is available here: \href{http://ecommons.library.cornell.edu/bitstream/1813/14112/2/ristenpart_gfm_2009_hires.mpg}{High resolution
version} or \href{http://ecommons.library.cornell.edu/bitstream/1813/14112/3/ristenpart_gfm_2009.mpg}{Low resolution version}.  The video
contains six segments, all displayed at 25 frames per second, and are arranged in order as follows.
\begin{enumerate}
\item This movie shows immediate coalescence at a low field strength.  A droplet of water (10 mM KCl, nominal diameter = 1.4 mm) moves through
silicone oil (PDMS, 1000 cS).   The black rectangular object at top is the tip of the electrode; the long black arc at bottom is the oil/water
meniscus (cf. Figure 1 in main text).  The applied potential was 0.5 kV, yielding an approximate field strength of 100 V/mm.  The capture frame
rate was 1000 frames/sec; the time lapse covers a period of 150 ms total.

\item This movie shows non-coalescence (bouncing) at a higher field strength.  The setup is identical to that shown in Movie 1,
except the applied potential was 1.0 kV (approximate field strength of 200 V/mm).  The capture frame rate was 1000 frames/sec; the time lapse
covers a period of 230 ms total.

\item This movie shows vigorous bouncing at an even higher field strength.  The setup is identical to that shown in  Movies 1 and 2,
 except the applied potential was 3.0 kV (approximate field strength of 600 V/mm).  The capture frame rate was 125 frames/sec; the time lapse covers
  a period of 4000 ms total.

\item This movie shows two droplets bouncing back and forth in a high field strength. The setup is identical to that shown in Movies
1-3, with the addition of a smaller second droplet (both drops 1 M KCl, with nominal diameters of 1.2 mm and 1 mm for top and bottom drops
respectively).  The applied potential was 3.0 kV (approximate field strength of 600 V/mm).  The capture frame rate was 500 frames/sec; the time
lapse covers a period of 2000 ms total.

\item This movie shows a zoomed-in view of the immediate vicinity at the point of contact for a bouncing drop.  The dark regions near the top and
bottom are water (1 M KCl) and the intervening bright region is oil (PDMS 1000 cS).  The total image width is 0.45 mm.   The lighter regions within
 the water are optical artifacts due to reflection. At the point of contact a clear meniscus bridge is observed; the corresponding still frame is
 shown in Fig. 3a in the main text.  The applied potential was 1.0 kV (approximate field strength of 200 V/mm).   The capture frame rate was 25,000
 frames/sec; the time lapse covers a period of 28 ms total.

\item This movie shows an example of partial coalescence, where an oppositely charged `daughter' droplet is emitted from the point of contact.  The
setup is similar to that shown in Supplementary Movie 3, except the viscosity of the silicone oil is decreased (100 cS) and the water salt
concentration is increased (1 M KCl).  The applied potential was 3.0 kV (approximate field strength of 600 V/mm).   The object at top-left is
the pipette tip.  The capture frame rate was 1000 frames/sec; the time lapse covers a period of 700 ms total.

\end{enumerate}
\end{document}